\documentstyle[twocolumn,aps,prb]{revtex}

\begin{document}
\draft

\title{Ballistic and diffuse transport through a  ferromagnetic domain wall}

\author{Arne Brataas$^{\dag}$}

\address{Department of Applied Physics and Delft Institute of
Microelectronics and Submicrontechnology, \\ Delft University of
Technology, Lorentzweg 1, 2628 CJ Delft, The Netherlands}

\author{Gen Tatara}

\address{Graduate School of Sciences, Osaka University, Toyonaka, Osaka
560,Japan}

\author{Gerrit E. W. Bauer}

\address{Department of Applied Physics and Delft Institute of
Microelectronics and Submicrontechnology, \\ Delft University of
Technology, Lorentzweg 1, 2628 CJ Delft, The Netherlands}

\date{\today} \maketitle

\begin{abstract}

We study transport through ballistic and diffuse ferromagnetic domain
walls in a two-band Stoner model with a rotating magnetization
direction.  For a ballistic domain wall, the change in the conductance
due to the domain wall scattering is obtained from an adiabatic
approximation valid when the length of the domain wall is much longer
than the Fermi wavelength.  In diffuse systems, the change in the
resistivity is calculated using a diagrammatic technique to the lowest
order in the domain wall scattering and taking into account
spin-dependent scattering lifetimes and screening of the domain wall
potential.

\end{abstract}


\section{Introduction}

In a ferromagnet, domains with different directions of the
magnetization are favored by the long-range magnetic dipolar
interaction. The boundary between the domains, the domain walls
(DW's), are a source of magnetoresistance that recently has attracted
experimental\cite
{Gregg96:1580,Hong98:L401,Otani98:1096,Ruediger98:5639,Gorkom99:422} and
theoretical interest.\cite{Tatara97:3773,Levy97:5110,vanHoof99:138}

For ballistic systems, where the electron mean free path is longer
than the system size, first-principles band-structure calculations
have shown that the DW resistance is enhanced due to the nearly
degenerate bands at the Fermi energy.\cite{vanHoof99:138} The rotating
magnetization direction causes an effective potential barrier for the
electrons which increase the resistance. Recently large DW resistance
in ballistic Ni nanocontacts has been measured\cite{Garcia99:2923} in
agreement with the predictions in Ref.\ \onlinecite{vanHoof99:138}.

In the diffuse transport regime Cabrera and
Falicov\cite{Cabrera74:217} interpreted transport through a single DW
as a tunneling process and the corresponding MR was found to be
exponentially small. Berger\cite{Berger84:1954} modelled the domain
wall scattering as a force on the magnetic moment of the conduction
electrons. Tatara and Fukuyama \cite{Tatara97:3773} calculated the DW
conductivity for spin-independent scattering life-times. Levy and
Zhang\cite{Levy97:5110} pointed out that spin-dependent impurity
scattering can strongly enhance the DW resistivity.

Beyond the semiclassical transport theories, Tatara and Fukuyama\cite
{Tatara97:3773} predict a negative DW resistivity as a result of the
reduced weak localization correction due to the decoherence of the
electrons by the scattering off the domain wall. However, quantum
interference effects do not explain the experiments in Refs.\
\onlinecite{Otani98:1096,Ruediger98:5639}, where the negative DW
resistance persists up to high temperature where the inelastic
scattering length is shorter than the mean free path.  It has been
suggested by Ruediger et al.\cite{Ruediger98} that the experimentally
observed negative DW resistance is an extrinsic effect caused by the
interplay between orbital effects due to the internal magnetic fields
and surface scattering. Recently, it was also demonstrated that the
large negative domain wall resistance of Co films\cite{Gregg96:1580}
is due to MR resistivity anisotropy.\cite{Ruediger9901245}

It is the purpose of the present paper to give a detailed account of
the transport through a domain wall both in the ballistic and the
diffuse regime in a two-band Stoner model. In the ballistic transport
regime, the transport through the magnetic domain wall can be treated
by an adiabatic approximation similar to the one used for transport
through a quantum point contact. In the diffuse regime we will use the
diagrammatic technique introduced in Ref.\ \onlinecite{Tatara97:3773}
and generalize it to the case of asymmetric impurity scattering
life-times (but without localization effects\cite{Tatara97:3773}) and
screening of the domain wall potential. Our results, although more
general, reduce in the case of strong spin splitting to results that
are very similar to those obtained in Ref.\ \onlinecite{Levy97:5110}
using a Boltzmann equation. We explain why the results of the two
methods differ. Some results have been published already in a brief
report and a conference proceedings.\cite{vanHoof99:138,Brataas98:545}
Here we give an in depth discussion of the results including the
technical details of the derivations.

The paper is organized in the following way. The two-band Stoner model
for the ferromagnet with a rotating magnetic field and how it can be
reduced to a more tractable form by a local gauge transformation is
discussed in section \ref{s:model}. The ballistic transport regime is
discussed in section \ref{s:ballistic} and the diffuse transport
regime in section \ref{s:diffuse}. We give our conclusions in section
\ref{s:conclusions}. The appendices include the adiabatic
approximation which can be used in the ballistic situation, recipes
for the calculation of the frequency summation of the Feynman
diagrams, and the spin-spiral case which can be exactly diagonalized.

\section{Model}
\label{s:model}

Throughout, we will use an effective 2-band model to describe the
ferromagnet with the Hamiltonian
\begin{equation}
H=\int d{\bf r}{\bf \Psi }_0^{\dagger }({\bf r})\left[ -\frac{\hbar
^{2}}{2m} \nabla ^{2}+\mu_{B}{\bf H}({\bf r})\cdot \text{\boldmath
$\sigma$}\right] {\bf \Psi }_0({\bf r}), \label{ham}
\end{equation}
where $\mu_{B}$ is the Bohr magneton, $\sigma _{x},\sigma _{y}$ and
$\sigma _{z}$ are the Pauli spin matrices, ${\bf H}({\bf r})$ is the
effective magnetic field arising from the electronic exchange
interaction, the magnetic dipole interaction, and the external
magnetic field and ${\bf \Psi}_0( {\bf r})$ is the 2-component spinor
wave function.  The direction of the effective magnetic field is
represented by a rotation angle $\theta (z)$ which varies along the
$z$-direction. The spin-orbit interaction and the Lorentz force due to
the internal magnetization are disregarded, since experimentally the
DW magnetoresistance can be separated from the anomalous
magnetoresistance (AMR) and the orbital magnetoresistance
(OMR).\cite{Ruediger98:5639} We use a local gauge
transformation\cite{Tatara97:3773} ${\bf \Psi}_0({\bf r}) = U({\bf r})
{\bf \Psi}({\bf r})$, where
\begin{equation}
U({\bf r}) = \cos(\theta/2)\sigma_z + \sin(\theta/2) \sigma_x
\label{gauge}
\end{equation}
and introduce the Fourier transform of the change of the direction of
the magnetic field $a(z) \equiv d\theta (z)/dz=\sum_{q}\exp
(iqz)a_q$. After the gauge transformation (\ref{gauge}) the
Hamiltonian (\ref{ham}) becomes $\widetilde{H}=U^{\dag} H U =
H_{0}+V$. The unperturbed term is
\begin{equation}
H_{0}=\sum_{{\bf k}s}\left( \epsilon _{{\bf k}}^s-\mu
\right) c_{{\bf k}s}^{\dagger }c_{{\bf k }s} \, ,
\end{equation}
where $s=+$ ($s=-$) denotes spin-up (spin-down) states, $\epsilon
_{{\bf k} }^s=\hbar ^{2}k^{2}/2m-s\Delta $ and the spin-splitting is
$\Delta =\mu_{B} | {\bf H}|$. The interaction with the DW is
\cite{Tatara97:3773}
\begin{eqnarray}
V & = &\frac{\hbar^2}{2m} \frac{1}{4} \sum_{{\bf k}qq's} a_{q-q'}
a_{q'} c_{{\bf k}+q{\bf z}s}^{\dag} c_{{\bf k}s} + \nonumber
\\ & & \frac{\hbar^2}{2m}\sum_{{\bf k}qss'}
\left(k_z+\frac{q}{2}\right) a_q c_{{\bf k} + q\hat{\bf z}s}^{\dag}
(\sigma_y)_{s,s'} c_{{\bf k}s'} \, ,
\label{V}
\end{eqnarray}
where $\hat{\bf z}$ is a unit vector in the $z$-direction and
$k_z={\bf k} \hat{\bf z}$.

\section{Ballistic transport}
\label{s:ballistic}

Transport through the domain wall is ballistic when the system size
is smaller than the mean free path. In this regime, the transport
properties can be described by the Landauer conductance,
\begin{equation}
G=\frac{e^2}{h} \sum_{{\bf k}_{\parallel}ss'} T_{{\bf
k}_{\parallel}}^{ss'} \, ,
\end{equation}
where $T_{{\bf k}_{\parallel}}^{ss'}$ is the transmission probability
for an electron in the transverse mode ${\bf k}_{\parallel}$ to pass
the domain wall from spin-state $s'$ to spin-state $s$. Domain walls
in transition metals are much thicker than the Fermi wavelength (the
length of the domain wall is $\lambda_w = 40$nm (Fe),
$\lambda_w=100$nm (Ni), $\lambda_w=15$nm (Co) and the Fermi wavelength
is roughly $\lambda_F \sim 0.2$ nm). The transmission probability can
therefore be calculated with the aid of an adiabatic approximation on
the eigenstates of the Hamiltonian after the gauge transformation
(\ref{gauge}) (see Appendix \ref{s:adiabatic}). The domain wall is
then equivalent to an effective potential barrier for the electrons.
The conductance is determined by the minimum number of propagating
modes, which is where the gradient of the rotating magnetic field has
its maximum value, $a(z) \equiv d \theta(z)/dz \rightarrow a_{\max}$
(for details see Appendix \ref{s:adiabatic}). The conductance can then
be found from the conductance of a spin-spiral with gradient
$a_{\max}$, which has the dispersion $E_{\max}^{\pm}=\hbar^2/(2m)
[k^2+a_{\max}^2/4\pm (k_z^2 a_{\max}^2+p^4)^{1/2}]$, and the
conductance is
\begin{equation}
G=\frac{e^2}{2} \sum_{{\bf k}s} |v_{s}| \delta(E_{\max}^{s}-E_F) \, ,
\end{equation}
where $v_{s} = \partial E_{s}/( \hbar \partial k_z)$ is the
group velocity.  Carrying out the integration, we find the domain wall
resistance $R_w/R_0=R-R_0$ ($1/R=G$, $1/R_0=G_0=(e^2A
\bar{k}_F^2)/(2\pi h)$):
\begin{equation}
\frac{R_w}{R_0} = \left\{ \begin{array}{ll} a_{\max}^2/(4\bar{k}_F^2) & \text{$a_{\max}^2 \le 2p^2$} \\ \ p^2/\bar{k}_F^2 - p^4/(a_{\max}^2 \bar{k}_F^2) &
 \text{$a_{\max}^2 > 2p^2$} \end{array} \right. \, ,
\end{equation}
where $\hbar^2 \bar{k}_{F}^2/(2m) \equiv E_F$ and $\Delta \equiv
\hbar^2 p^2/(2m)$.  The screening of the domain wall potential
discussed below is not important for the calculation of the
conductance, since by a calculation following the lines in section
\ref{s:diffuse}, we find that the shift in the chemical potential due
to the rotating magnetization is $\delta \mu \approx - (1/48) E_w
(\Delta/E_F)^2 + {\cal O}(\Delta/E_F)^3$. The conductance is $G \sim
k_F^2 \sim \mu$, and therefore screening gives a vanishing small
contribution to the change in the resistance when the splitting is
sufficiently small, $\delta R_{\mu}/R_0 \approx (1/48) (E_w/E_F)
(\Delta/E_F)^2$.

Using parameters for Fe, Ni and Co ($\lambda_w=40$ nm, $\lambda_w=100$
nm, $\lambda_w=15$ nm, respectively), we find $R_w/R=0.0008 \%$,
$R_w/R_0 = 0.0001 \%$, $R_w/R_0= 0.008 \%$ respectively. Within the
2-band model, the ballistic domain wall scattering is thus very
weak. In first-principles band structure calculations these small
numbers are enhanced by orders of magnitude due to the (near)
degeneracy of the energy bands at the Fermi level.\cite{vanHoof99:138} 

\section{Diffuse transport}
\label{s:diffuse}

When the system size is much larger than the mean free path, the
transport is in the diffuse regime.  We assume that the electrons are
subject to spin-dependent scattering, which is modelled by short-range
scatters giving rise to spin-dependent life-times $\tau_+$ and
$\tau_-$ for spin-up and spin-down states, respectively, which will be
treated as adjustable parameters.

We study the current in the $z$-direction. The current operator
transformed by the local spin rotation $U({\bf r})$ in (\ref{gauge})
is $\tilde{J}=U^{\dag} J U=J_{0}+J_{g}$. The unperturbed current
operator is
\begin{equation}
J_{0}=\frac{e\hbar}{m} \sum_{{\bf k}s }k_{z}c_{{\bf k}s }^{\dagger
}c_{ {\bf k}s}
\label{cur}
\end{equation}
and due to the local gauge transformation\cite{Tatara97:3773}
\begin{equation}
J_g = \frac{e\hbar}{2m}\sum_{{\bf k} q ss'} a_q c_{{\bf k}
+ q{\bf z}s} (\sigma_y)_{s,s'} c_{{\bf k}s'} \, .
\label{gaugecur}
\end{equation}
The conductivity is calculated from the Kubo formula. 
\begin{equation}
\sigma(\omega) = \frac{i}{\omega} \left[\Pi(\omega) + \frac{n_0e^2}{m} \right] \, ,
\end{equation}
where $n_0=N/V$ is the electron density, $N$ is the number of
electrons and $V$ is the volume of the system. The current-current
correlation function $\Pi(\omega)$ is obtained by an analytical
continuation ($i\omega_l \rightarrow \omega + i\delta$, $\delta
\rightarrow 0^+$) from the Matsubara correlation function
\begin{equation}
\Pi^l = - \frac{1}{V}\frac{1}{\hbar} \int_0^{\beta \hbar} e^{i\omega_l
t} \langle |T_{\tau} \tilde{J}(\tau) \tilde{J}(0) | \rangle \, ,
\end{equation}
where $1/\beta=k_B T$, $k_B$ is the Boltzmann constant, $T$ is the
temperature and $T_{\tau}$ is the $\tau$-ordering operator.  We will
only study the DC conductivity at low temperatures by letting $\omega
\rightarrow 0$ and $T \rightarrow 0$.  The relevant Feynman diagrams
to the lowest order in the scattering by the domain wall were
identified in Ref.\ \onlinecite{Tatara97:3773} and are shown in
Fig.\ \ref{Fdiagram}. Diagram (0) represents the zeroth order Drude
contribution, diagram (1) is due to the correlation of the correction
to the conductivity operator (\ref{gaugecur}), diagrams (2) and (4)
are self-energy corrections from the interaction Hamiltonian (\ref{V})
to the electron Green's function, diagram (5) is a vertex correction,
and diagram (3) is the correlation of the change in the current
operator (\ref{gaugecur}) and the interaction Hamiltonian
(\ref{V}). The electron Green's function appearing in the Feynman
diagrams in Fig.\ \ref{Fdiagram} is a configurational average over
impurity positions, {\it e.g.} the retarded Green's function is
\begin{equation}
G_{{\bf k}s}^{R}(\omega) = \frac{1}{\hbar \omega - \epsilon_{\bf k}^s
+ i \hbar/2\tau_{s}} \, .
\end{equation}
The scattering lifetimes of the states at the Fermi energy due to the
impurity scattering $\tau _{s}$ are assumed to be isotropic but
may be spin-dependent.\cite{Mertig94:11767}

The DC conductivity of a single-domain ferromagnet is 
\begin{eqnarray}
\sigma _{0}& =& \frac{e^{2}}{V}\sum_{{\bf k}s}\left( \frac{\partial
\epsilon _{{\bf k}s}}{\partial k_{z}}\right) ^{2}\delta (\epsilon
_{{\bf k} s}-\mu_0)\tau _{s} 
\label{Ddetail}\\
& =&\frac{e^2}{m}\left( n_{+}
\tau_{+} + n_{-} \tau_{-}\right),
\label{Drude}
\end{eqnarray}
where $\mu_0$ is the bulk chemical potential, $n_{+}$ ($n_{-}$) is the
electron density of spin-up (spin-down) states and $\tau_{+}$
($\tau_{-}$) is the scattering life-time of spin-up (spin-down)
states. 

There are two contributions to the conductivity which to the lowest
order in the domain wall scattering are additive. Firstly, screening
shifts the chemical potential and induces a DW conductivity in
(\ref{Ddetail}). Secondly, the electrons are directly scattered by the
domain wall by the interaction term (\ref{V}) and the gauge
transformed current operator (\ref{gaugecur}).

Since the width of domain walls in transition metals is much larger
than the screening length, electroneutrality dictates that the
electron density to a good approximation is the same in the presence
or absence of the DW, but the chemical potential differs. This is in
contrast to the treatment in Refs.\
\onlinecite{Tatara97:3773,Levy97:5110}, where the chemical potential
is assumed to be the same in the presence or absence of the DW.  The
change in the conductivity due to the chemical potential shift can be
found from (\ref{Ddetail}) setting $\mu_0 \rightarrow \mu_0 + \delta
\mu$:
\begin{equation}
\delta \sigma _{0}= \delta \mu \frac{e^{2}}{m} \sum_{s}N_{s}\tau _{s}
\label{sigshift} \, ,
\end{equation}
which to lowest order in the domain wall scattering may be added to
the DW conductivity. Here $N_{s}=mk_{F}^s/(2\pi ^{2}\hbar ^{2})$ is
the electron density of states at the Fermi energy, $k_{F}^s$ is the
spin-dependent Fermi wave vector related to the spin-dependent
electron density by $n_{s}=(k_{F}^s)^3/(6\pi^2)$ and we also introduce
$\epsilon_{F}^{s} \equiv \hbar^2 (k_{F}^s)^2/(2m)$.

We proceed by calculating the chemical potential shift due to the
rotating magnetization. The zeroth order contribution to the electron
density is $ n_{0}=\sum_{{\bf k}s}\theta (\mu -\epsilon _{{\bf
k}}^{s})/V,$ where $\theta (x) $ is the Heaviside step-function. The
Feynman diagrams of the contributions to the electron density in the
lowest order interaction with the domain wall are shown in Fig.\
\ref{Fdensity}. Diagram (A) is due to the first term in (\ref{V}) and
diagram (B) is due to the second order contribution of the second term
in (\ref{V}).  Combining the two terms, the second order contribution
to the electron density is
\begin{equation}
n_{2}=\frac{\hbar ^{2}}{2mV}\sum_{{\bf k}qs}\left| a_{q}\right|
^{2}\left( \frac{\hbar ^{2}k_{z}^{2}/2m}{2\Delta ({\bf k}q)}-\frac{1}{4}
\right) \delta (\epsilon _{{\bf k}_{-}}^{s}-\mu ),  \label{n2}
\end{equation}
where $2\Delta ({\bf k} q)=\epsilon _{{\bf k}_{+}}^{-s}-\epsilon _{{\bf
k} _{-}}^{s}$ and ${\bf k}_{\pm }={\bf k}\pm \left( q/2\right) {\bf
\hat{z}}$. Since the DW is much thicker than the Fermi wavelength, we
disregard the wave-vector dependence ($q$) on the electron states at
the Fermi level and introduce the energy parameter for the domain
wall, $E_w=\sum_q \hbar^2 |a_q|^2/(2m)$, {\it e.g.} with $\cos \theta =\tanh
(z/\lambda _{W})$, $E_{w}=\pi \hbar ^{2}/(L\lambda _{W}m)$, where
$n_W=1/L$ is the ``density'' of the domain wall.  The chemical potential shift
follows from $n_{0}+n_{2}=n,$ where $n$ is the electron density,
\begin{equation}
\delta \mu = E_w \left[ \frac{1}{4} - \frac{\sum_{s} s
N_{s} \epsilon_{F}^{s}}{6\Delta \sum_{s}
N_{s}}\right] \, .
\label{dmu}
\end{equation}
The correction in the conductivity (\ref{sigshift}) due to the
chemical potential shift becomes
\begin{equation}
\delta \sigma_0 = \frac{e^2E_w}{m}\sum_{s} N_{s} \tau_{s}
\left[\frac{1}{4} -\frac{\sum_{s}s N_{s}\epsilon
_{F}^{s}}{6\Delta \sum_{s}N_{s}} \right] \, .
\end{equation}

The corrections to the current-current correlation function from
diagrams (1--5) are
\begin{mathletters}
\begin{eqnarray}
\Pi_1^l & = & \frac{e^2\hbar^2}{4m^2V} \sum_{{\bf k}qs} |a_q|^2
\pi_1^l ({\bf k}qs) \\
\Pi_2^l & = & \frac{e^2\hbar^4}{8m^3V} \sum_{{\bf k}qs} |a_q|^2 k_z^2
\pi_2^l({\bf k}qs) \\
\Pi_3^l & = & \frac{e^2\hbar^4}{2m^3V} \sum_{{\bf k}qs} |a_q|^2
(k_z-\frac{q}{2})k_z \pi_3^l({\bf k}qs)\\
\Pi_4^l & = & \frac{e^2\hbar^6}{4m^4V} \sum_{{\bf k} qs} |a_q|^2
(k_z-\frac{q}{2})^2 k_z^2 \pi_4^l({\bf k}qs)\\
\Pi_5^l & = & \frac{e^2\hbar^6}{4m^4V} \sum_{{\bf k}qs} |a_q|^2 k_z^2
(k_z^2-\frac{q^2}{4}) \pi_5^l({\bf k}qs) \, ,
\end{eqnarray} 
\end{mathletters}
where the frequency summations are defined by
\begin{mathletters}
\begin{eqnarray}
\pi_1^l & = & \frac{1}{\beta} \sum_n G^{n+l}_{{\bf
k}_-s} G_{{\bf k}_+ -s}^n\\ 
\pi_2^l & = & \frac{1}{\beta} \sum_n \left[ G^{n+l}_{{\bf k}s} G^{n+l}_{{\bf k} s} G^{n}_{{\bf k}s} + (l \rightarrow -l) \right]\\ 
\pi_3^l & = & \frac{1}{\beta}\sum_n \left[G^{n+l}_{{\bf k}_-s} G^{n+l}_{{\bf k}_+ -s} G^{n}_{{\bf k}_-s} + (l \rightarrow -l) \right] \\ 
\pi_4^l & = & \frac{1}{\beta}\sum_n \left[ G^{n+l}_{{\bf k}_-s} G^{n+l}_{{\bf k}_+ -s} G^{n+l}_{{\bf k}_- s} G^{n}_{{\bf k}_- s} + (l \rightarrow -l) \right]\\ 
\pi_5^l & = & \frac{1}{\beta} \sum_n G^{n+l}_{{\bf k}_-s}
G^{n+l}_{{\bf k}_+-s} G^{n}_{{\bf k}_-s}
G^{n}_{{\bf k}_+ - s} \, .
\end{eqnarray}
\label{sums}
\end{mathletters}
In the low impurity density limit the energy splitting between the
bands is larger than the broadening of the bands due to the impurity
scattering, $\Delta \tau/\hbar \gg 1$.  The frequency sums
(\ref{sums}) are evaluated in Appendix \ref{s:freq}, where the weak
wave-vector dependence on the electron states at the Fermi level is
disregarded consistent with the treatment of the chemical potential
shift above. Carrying out the Matsubara frequency sums over the
internal energies at low temperature, we find that the correction to
the DC conductivity due to the DW is
\begin{mathletters}
\begin{eqnarray}
\label{sigma1}
\sigma_{1} & = & 0\\ 
\sigma_{2} &=&-\frac{e^{2}E_w}{m}\sum_{s}N_{s}\tau
_{s} \frac{1}{4} \\ 
\sigma_{3}
&=&-\frac{e^{2}E_w}{m}\sum_{s}N_{s}\tau _{s}
\frac{2}{3}s\frac{\epsilon _{F}^s}{\Delta } \\ 
\sigma _{4} &=&\frac{e^{2}E_w}{m}\sum_{s} N_{s}
\tau_{s}\left[\frac{s \epsilon _{F}^s}{2\Delta
}-\left( \frac{ \epsilon _{F}^s}{\Delta }\right) ^{2}\frac{\tau_{+}+\tau _{-}}{10 \tau _{-s}} \right] \\
\sigma _{5} &=&\frac{e^{2}E_w}{m}\sum_{s}N_{s}\tau
_{s} \frac{1}{5}\left( \frac{\epsilon _{F}^s}{\Delta
}\right) ^{2} \, .
\end{eqnarray}
\end{mathletters}

The contribution to the conductivity from the diagrams (1-5) in Fig.\
\ref{Fdiagram} is
\begin{eqnarray}
\sum_{i=1}^5 \sigma_{i} &=&-\frac{e^{2}E_{w}}{m}\sum_{s}N_{s}\tau _{s}\times  \nonumber \\
&&\left[ \frac{1}{4}+\frac{1}{6}s\frac{\epsilon _{F}^{s}}{\Delta 
}-\frac{1}{10}\left( \frac{\epsilon _{F}^{s}}{\Delta }\right)
^{2}\left( 1-\frac{\tau _{s}}{\tau _{-s}}\right) \right] \, .
\end{eqnarray}
The DW resistivity can be found from $\rho _{w}=-\sigma _{w}\rho
_{0}^{2}$ , where the DW conductivity change due to the rotating
magnetic field is $ \sigma_{w}=\delta \sigma_{0}+\sum_{i=1}^5 \sigma
_{i} $:
\begin{eqnarray}
\rho_{w} &=&\frac{e^{2}\rho^2_{0}E_{w}}{6m}\sum_{s}N_{s}\tau _{s}\times  \nonumber \\
&&\left[ \frac{\delta \epsilon _{F}}{\Delta }+\frac{s\epsilon
_{F}^{s}}{\Delta }-\frac{3}{5}\left( \frac{\epsilon _{F}^{s}}{
\Delta }\right) ^{2}\left( 1-\frac{\tau _{s}}{\tau _{-s}}\right) 
\right]  \label{rhoR} \, ,
\end{eqnarray}
where $\delta \epsilon _{F}=\sum_{s}s N_{s}\epsilon
_{F}^{s}/\sum_{s}N_{s}$. The first term in (\ref{rhoR}) is always
positive, but the second and the third terms can be negative when the
relaxation time of the minority spin electrons is longer than the
relaxation time for the majority spin electrons. However, as will be
demonstrated below, the domain wall resistivity given by (\ref{rhoR})
is always positive. Our speculation about the possibility of a
negative domain wall resistance in Ref.\ \onlinecite{Brataas98:545} is
thus not justified from Eq.\ (\ref{rhoR}) only. The result
(\ref{rhoR}) differs from that obtained in Ref.\
\onlinecite{Tatara97:3773} for spin-independent relaxation times $\tau
_{s}=\tau $, where screening was not taken into account, {\it i.e.} a
constant chemical potential and not a constant electron density was
assumed. The result also differs from the calculation in Ref.\
\onlinecite{Levy97:5110} based on the Boltzmann equation. We believe
that this latter discrepancy is because in Ref.\
\onlinecite{Levy97:5110} screening as well as the effect of the gauge
transformation on the current operator (\ref{gaugecur}) are
neglected. The latter corresponds to the neglect of the diagrams (1)
and (3) in Fig.\ \ref{Fdiagram}. For sufficiently weak impurity
scattering and/or a large spin splitting ($\Delta \tau/\hbar \gg1$)
the contribution from diagram (1), Eq.\ (\ref{sigma1}), vanishes, so
only in this limit the omission of this diagram is
justified. Furthermore, the difference in the Fermi wave vectors for
the spin-up and spin-down electrons was disregarded in parts of the
calculations in Ref.\ \onlinecite{Levy97:5110} and an approximation
was introduced in order to solve the integral equation for the
Boltzmann equation.  Indeed, assuming small spin splittings ($\Delta /
\epsilon_F \ll 1$, but $\Delta\tau/\hbar \gg 1$) in Eq.\ (\ref{rhoR}),
we obtain 
\begin{equation}
\frac{\rho_w}{\rho_0} \approx \frac{3E_w E_F}{20\Delta^2}
\frac{(\tau_+-\tau_-)^2}{\tau_+ \tau_-}
\end{equation}
 which is very similar, but not idential to the result in Ref.\
\onlinecite{Levy97:5110}. In this limit, the domain wall resistivity
increases quadratically with the asymmetry in the spin-up and
spin-down scattering lifetimes as pointed out in Ref.\
\onlinecite{Levy97:5110}. For larger spin splitting, Eq.\ (\ref{rhoR})
should be used.

It is also interesting to study the domain wall resistivity when the
spin-splitting is large, $2\Delta > \mu$, which is the case for a
half-metallic ferromagnet in which the minority spin density of
states vanishes $N_-=0$. In this regime we find
\begin{equation}
\frac{\rho_w}{\rho_0} = \frac{E_w}{\mu} \left[\frac{\mu}{2\Delta} -
\frac{3}{5} (\frac{\mu}{2\Delta})^2 (1-\frac{\tau_+}{\tau_-}) \right]
\, .
\label{rho1band}
\end{equation}
The first term in (\ref{rho1band}) can be interpreted as additional
intraband scattering in the majority spin channel due to the rotating
magnetization and the second term as virtual transport in the minority
spin channel which has a negative contribution when
$\tau_->\tau_+$. The domain wall resistivity (\ref{rho1band}) is
always positive. In the limit of large spin-splittings $\Delta \gg
\mu$ the domain wall resistivity vanishes, since the coupling between
the bands becomes vanishing small. Note that the present formalism is
valid only for wide walls, since the domain wall scattering is treated
as a perturbation.

Our perturbative result (\ref{rhoR}) can be checked against an exact
calculation for a spin-spiral, $d \theta(z) /dz=a_0$,
$a_{q}=a_{0}\delta _{q,0}$ with spin-independent life-times $\tau_s
\rightarrow \tau$. The detailed calculation is shown in Appendix
\ref{s:spinspiral}. The Hamilonian is diagonalized in spin space by
${\bf u} _{\pm }={\cal N}_{\pm }[1,i(1\mp \sqrt{1+\alpha ^{2}}
)/\alpha ]^{T}$, where ${\cal N}_{\pm }$ is a normalization constant,
$ \alpha =k_{z}a_{0}/p^{2}$ and $\Delta =\hbar ^{2}p^{2}/2m$. The
corresponding eigenvalues are $E_{\bf k}^{\pm }=\left( \hbar ^{2}/2m\right)
\left( k^{2}+a_{0}^{2}\mp \sqrt{k_{z}^{2}a_{0}^{2}+p^{4}}\right)
$. The (Drude) resistivity can be calculated from the Kubo formula,
\begin{equation}
\rho_{w}=\frac{e^{2}\rho _{0}^{2}E_{w}}{2m\Delta }\tau (n_{+}-n_{-})\,.
\label{exact}
\end{equation}
This is in exact agreement with Eq.\ (\ref{rhoR}) for $a_q=a_0
\delta_{q,0}$ when $\tau_+=\tau_-$; a good indication of the
correctness of our perturbation approach. The calculations in Refs.\
\onlinecite{Tatara97:3773,Levy97:5110} disagree with the exact result
(\ref{exact}) presumably due to the reasons outlined above.

The result for the domain wall resistivity (\ref{rhoR}) can be
analyzed by introducing $k_{F}^{+} = \sqrt{\gamma} k_{F}$,
$k_{F}^{-}=k_{F}/\sqrt{\gamma}$, $\tau_{+}= \sqrt{\eta} \tau$,
$\tau_-=\tau / \sqrt{\eta}$, where $\gamma=k_{F}^{+}/k_{F}^-$ is a
measure of the polarization of the ferromagnet, $\eta=\tau_+/\tau_-$
is a measure of the asymmetry of the scattering lifetimes, $k_F$ is
the average Fermi wave vector and $\tau$ is the average scattering
life-time. The domain wall resistivity is proportional to
$\kappa=E_W/4E_F$.  Typically in Fe $k_F \sim 1.7\AA$, $\lambda_W \sim
300 \AA$ and $n_w \sim 2.5 \mu m^{-1}$ giving $\kappa\sim 10^{-6}$,
which means that the domain wall scattering for symmetric scattering
life-times is very weak. However, as pointed out in Ref.\
\onlinecite{Levy97:5110}, the domain wall resistivity can become
appreciable larger when taking into account the life-time asymmetry of
the carriers. We show in Fig.\ \ref{f:rho} the scaled domain wall
resistivity $\rho_w/(\kappa \rho_0)$ as a function of the asymmetric
scattering life-times $\tau_+/\tau_-$ in the case of a small spin
polarization $\gamma=1.01$ (solid line), intermediate spin
polarization $\gamma=1.20$ (dashed line) and large spin polarization
$\gamma=10.0$ (dotted line). For a larger spin-polarization, the domain
wall resistivity naturally becomes asymmetric in the relative
difference in the scattering lifetimes $\tau_+$ and $\tau_-$.  The
domain wall resisitivity becomes noticable for asymmetric life-times
and can become of the order $\rho_w/\rho_0 \sim 1\%$.

\section{Conclusions}
\label{s:conclusions}

We studied the contribution of domain wall scattering on the transport
properties of a ferromagnet using an effective 2-band model.

In a diffuse ferromagnet, the domain wall resistivity is calculated
from the Kubo formula. The domain wall resistivity is found to be
strongly enhanced when the scattering lifetimes of the majority spins
and minority spins are different, in agreement with the results in
Ref.\ \onlinecite{Levy97:5110}.

In the ballistic regime, we have demonstrated how the domain wall
scattering creates an effective barrier that the electrons must
pass. The results from the 2-band model give only very small
corrections to the resistance of the system.

First-principle band-structure calculations have shown that the domain
wall resistance can be increased by orders of magnitude in the
ballistic regime.\cite{vanHoof99:138} It would be interesting to
perform a realistic band-structure calculation also for diffuse
systems. However, the generalization of our 2-band results turns out
to be cumbersome.\cite{Brataas98:545} 

\acknowledgements

This work is part of the research program for the ``Stichting voor
Fundamenteel Onderzoek der Materie'' (FOM), which is financially
supported by the ''Nederlandse Organisatie voor Wetenschappelijk
Onderzoek'' (NWO). We acknowledge benefits from the TMR Research
Network on ``Interface Magnetism'' under contract No. FMRX-CT96-0089
(DG12-MIHT) and support from the NEDO joint research program
(NTDP-98). We acknowledge stimulating discussion with Jaap Caro, Ramon
P. van Gorkom, Junichiro Inoue, Paul J. Kelly, and Andrew D. Kent.

\appendix

\section{Adiabatic approximation}
\label{s:adiabatic}

In the ballistic regime, it is most convenient to start with the
Hamiltonian in its first quantized form, which after the gauge
transformation (\ref{gauge}) reads
$\tilde{H}=H_0+V$:\cite{Levy97:5110}
\begin{eqnarray}
H_0 & = & -\frac{\hbar^2}{2m} \nabla^2 +\Delta \sigma_z \, ,\\
V & = & \frac{\hbar}{2m} \sigma_y a(z)p_z - \frac{i\hbar^2}{2m}
\sigma_y a'(z) + \frac{\hbar^2}{8m} a^2(z) \, ,
\end{eqnarray}
where $a(z)=d \theta(z)/dz$ is the gradient of the rotating angle of the
magnetization and $a'(z)=d^2 \theta(z)/dz^2$.  The wave function can be
written as
\begin{equation}
\Psi({\bf r}) = \phi(\text{\boldmath $\rho$}) \left[A_{\bf k}(z)
\left( \begin{array}{c} 1 \\ 0 \end{array} \right) + B_{\bf k}(z)
\left( \begin{array}{c} 0 \\ 1 \end{array} \right) \right] \, ,
\end{equation}
where $\phi(\text{\boldmath $\rho$})=A^{-1/2}\exp(i{\bf k}_{\parallel}
\text{\boldmath $\rho$})$ is the transverse part of the wave function
(${\bf k}=({\bf k}_{\parallel},k_z)$), $A_{\bf k}(z)$ is the spin-up
like longitudinal amplitude and $B_{\bf k}(z)$ is the spin-down like
longitudinal amplitude. The Schr\"{o}dinger equation then becomes
\begin{equation}
\left(
\begin{array}{cc}
-\frac{d^2}{dz^2}-\omega_+^0 & -a \frac{d}{dz} - \frac{a'}{2} \\
a+\frac{a'}{2} & -\frac{d^2}{dz^2} -\omega_-^0
\end{array}
\right) 
\cdot
\left(
\begin{array}{c}
A_{\bf k} \\
B_{\bf k}
\end{array}
\right)
=
\left(
\begin{array}{c}
0 \\
0
\end{array}
\right) \, ,
\label{Hadia}
\end{equation}
where 
\begin{equation}
\omega_{\pm}^0 =k_{\perp}^2-a(z)^2/4\mp p^2 \, ,
\end{equation}
$k_{\perp}^2=2mE_F/\hbar^2-k_{\parallel}^2$, and $E_F$ is the Fermi
energy. The off-diagonal terms in (\ref{Hadia}) describe the coupling
between the spin-up and spin-down like states. In the case of a
spin-spiral, $a'(z)=0$, the eigenstates can be found to be $A_{\bf k} =
A_{\bf k}^0 \exp(ik_z z)$, $B_{\bf k} = B_{\bf k}^0 \exp(i k_z z)$,
where the dispersion of the modes, $k_z$, are determined by
\begin{equation}
k^2_{\perp}=k_z^2+a^2/4\pm \sqrt{p^4 + q^2 k_z^2} \, ,
\end{equation}
The coupling is weak when the gradient of the spin rotation gradient
is slow compared to the Fermi wave length.  This permits us to make
use of a multiple scale analysis (or adiabatic
approximation).\cite{Bender78} This analysis is done by introducing
the small parameter $\epsilon$, so that $\omega(z) \rightarrow
\omega(\epsilon z)$, $a(z) \rightarrow a(\epsilon z)$ and $da/dz
\rightarrow \epsilon da/dz$ and we introduce the new variable
$Z=g(\epsilon z)/\epsilon$, where $g(\epsilon z)$ is a scaling
function.\cite{Bender78} We expand the longitudinal function $A_{\bf
k}(z)$ and $B_{\bf k}(z)$ to the lowest order in the small parameter
$\epsilon$, $A_{\bf k}(z,Z)=a_{\bf k}^0(Z,z)+ \cal{O}(\epsilon)$. To
the lowest order in the small parameter $\epsilon$, the equation to
solve is thus
\begin{equation}
\left(
\begin{array}{cc}
\frac{d^2}{dZ^2} + \frac{\omega_+^0}{(g')^2} & \frac{a}{g'} \\
-\frac{a}{g'} & \frac{d^2}{dZ^2} +\frac{\omega_-^0}{(g')^2} 
\end{array}
\right)
\cdot
\left(
\begin{array}{c}
a_{\bf k}^0 \\
b_{\bf k}^0
\end{array}
\right)
=
\left(
\begin{array}{c}
0 \\
0
\end{array}
\right) 
\, .
\end{equation}
We now make the ansatz $a^0_{\bf k}(z,Z)=a^{0,0}_{\bf k}(z) \exp(i Z)$
and find that the scaling function $g$ is chosen such that
\begin{equation}
\left( (g')^2-\omega_- \right) \left((g')^2-\omega_+\right)=p^4+(ag')^2
\label{scaling}
\end{equation}
and $Z=\frac{1}{\epsilon} \int^z dx g'(x)$, so that the adiabatic
solution is
\begin{equation}
A_{\bf k}(z) \sim a_0(z) \exp \left(\int^z dx g'(x) \right) \, .
\end{equation}
Similarly we can find a solution for $B_{\bf k}(z)$. Disregarding
tunneling states which only give an exponentially small contribution
to the conductance, the number of propagating modes is determined by
the condition $\text{Im}[ g'(x)]=0$. From (\ref{scaling}), we see that
the number of propagating modes is determined by the position where
$a(z)$ attains its maximum, {\it i.e.} the conductance can be
calculated as for a spin-spiral with $a(z) \rightarrow a_{\max}$.

\section{Frequency summations}
\label{s:freq}

The typical frequency sum to be perfomed is 
\begin{equation}
\pi^l=\frac{1}{\beta }\sum_{n} X^{n+l}Y^n+ (l \rightarrow -l) ,
\label{freqsum}
\end{equation}
where $X^n$ and $Y^n$ are Matsubara Green's
functions. They can be written in the spectral representation
\begin{equation}
X^n=\int_{-\infty }^{\infty }\frac{d\epsilon }{2\pi }\frac{
S_{X}(\epsilon )}{i\omega _{n}-\epsilon },
\end{equation}
where the spectral function is determined by the retarded and the
advanced function
\begin{equation}
S_{X}(\epsilon )=i\left[ X^{R}(\epsilon )-X^{A}(\epsilon )\right] \, .
\end{equation}
Performing the frequency summation in (\ref{freqsum}), we get
\begin{eqnarray}
\pi^l & = &\int_{-\infty }^{\infty }\frac{d\epsilon _{1}}{2\pi }
\int_{-\infty }^{\infty }\frac{d\epsilon _{2}}{2\pi }S_{X}(\epsilon
_{1})S_{Y}(\epsilon _{2}) \times \nonumber \\
&& \left[ \frac{f(\epsilon _{2})-f(\epsilon
_{1})}{ i\omega _{l}-(\epsilon _{1}-\epsilon _{2})}-\frac{f(\epsilon
_{1})-f(\epsilon _{2})}{i\omega _{l}-(\epsilon _{2}-\epsilon
_{1})}\right] \, .
\end{eqnarray}
The DC-conductivy is obtained by an analytical continuation 
\begin{equation}
I \equiv - \lim_{\omega \rightarrow 0} \frac{\pi^l(i\omega _{l}\rightarrow \omega
+i\delta)}{\omega }
\end{equation} 
and we consider the limit of zero temperature ($T\rightarrow 0$):
\begin{eqnarray}
I&=&\frac{\hbar }{ 2\pi }S_{X}(0)S_{Y}(0) \\ &=&\frac{\hbar }{\pi
}\mbox{Re}\left[ X^{R}(0)Y^{A}(0)-X^{R}(0)Y^{R}(0) \right] \, .
\end{eqnarray}
The product of the two retarded (advanced) Green's functions vanishes
when integrating over the energy since the poles are on the same side
of the imaginary plane. The sum can then be simplified to
\begin{equation}
I=\frac{\hbar }{\pi 
}\mbox{Re}\left[ X^{R}(0)Y^{A}(0)\right] \, .
\end{equation}
This relation will be used in the following in order to calculate the
contributions from the diagrams 1--5.

We use  
\begin{equation}
G_{s'}^{R}(0)G_{s}^{A}(0)=\frac{G_{s'}^{R}(0)-G_{s}^{A}(0)}{-i(\delta
_{s}+\delta _{s'})-(\epsilon^{s}-\epsilon^{s'})} \, ,
\end{equation}
where $\delta_s=\hbar /(2\tau_s)$ and obtain in the limit $\delta_s
\ll \mu$ the contribution from $\pi_1$ to the conductivity
\begin{equation}
I_{1} \approx [1+( \frac{\epsilon^{s}-\epsilon^{s' }} { \delta
_{s}+\delta _{s'}} ) ^{2}]^{-1} \frac{\tau _{s}\tau _{s'}}{\tau
_{s}+\tau _{s'}}\left[ \delta (\xi ^{s'})+\delta (\xi^{s})\right] \, ,
\end{equation}
where $\xi^s=\epsilon^{s}-\mu$ is the quasiparticle energy relative to
the Fermi level. In the case of no spin-splitting ($\xi^{s}\rightarrow
\xi $) and ($\delta _{s} \rightarrow \delta$), the result is $
I_{1}\approx \tau \delta (\xi ) $. In the limit of strong
spin-splitting, the result is vanishing small (of order $\hbar /\Delta
\tau $ small).
\begin{equation}
I_1 \approx 0 \, .
\end{equation}

The sum $\pi^{2}$ gives a contribution

\begin{equation}
I_{2} = \frac{\hbar }{2\pi }\frac{\partial }{\partial
\xi^{s}}\frac{1}{(\xi^{s})^{2}+\delta _{s}^{2}} \approx \tau
_{s}\frac{\partial }{\partial \xi^{s}}\delta (\xi^{s}) \, .
\end{equation}

  The sum $\pi^{3}$ gives a contribution 
\begin{equation}
I_{3} =-\frac{\hbar }{\pi }\frac{1}{(\xi^{s})^{2}+\delta
_{s}^{2}}\frac{ \xi^{-s}}{(\xi^{-s})^{2}+\delta _{-s}^{2}} \, .
\end{equation}
In the limit of vanishing spin-splitting and equal life-times, the
contribution is $ I_{3}\approx \tau (\partial /\partial \xi
)\delta (\xi )$, which agrees with the result of $\pi_{2}$ as it
should. When $\Delta \tau _{s}/\hbar \gg 1$ (large spin-splitting) it
is
\begin{equation}
I_{3}\approx -\frac{2\tau _{s}}{\epsilon^{-s}-\epsilon^{s}}\delta
(\xi^{s}) \label{q3} \, .
\end{equation}

In the case of no spin-splitting, the sum $\pi^4$ gives a contribution
\begin{equation}
I_{4}\approx \hbar \left[ -\frac{1}{4}\delta (\xi )\delta
^{-3}+\frac{1}{8}\delta ^{\prime \prime }(\xi )\delta ^{-1}\right] \,
.
\end{equation}
In the general case, we use
\begin{equation}
G_{s}^{R}(0)G_{-s}^{R}(0)=\frac{G_{s}^{R}(0)-G_{-s}^{R}(0)}{i\left(
\delta _{-s}-\delta _{s}\right) -\left(
\epsilon^{-s}-\epsilon^{s}\right) } \, .
\end{equation}
For large splitting the result is then 
\begin{eqnarray}
I_{4} & \approx & \frac{-\tau _{s}}{\epsilon^{-s}-\epsilon^{s} }\delta
^{\prime }(\xi^{s})-\frac{2\tau _{s}}{\left(
\epsilon^{-s}-\epsilon^{s}\right) ^{2}}\delta (\xi^{s})+ \nonumber \\
&& \frac{ \tau _{s}\left( 1-\frac{\tau _{s}}{\tau _{-s}}\right) }{
\left( \epsilon^{-s}-\epsilon^{s}\right) ^{2}}\delta (\xi^{s})
\label{q4} \, .
\end{eqnarray}

Finally, the sum $\pi_{5}$ gives a contribution
\begin{equation}
I_{5} = \frac{\hbar }{2\pi }\frac{1}{(\xi^{s})^{2}+\delta
_{s}^{2}}\frac{1}{(\xi^{-s})^{2}+\delta _{-s}^{2}} \, .
\end{equation}
In the case of no spin-splitting the sum is 
\begin{equation}
I_{5} \approx \hbar \left[ \frac{1}{4}\delta (\xi )\delta
^{-3}+\frac{1}{8}\delta ^{\prime \prime }(\xi )\delta ^{-1}\right] \, .
\end{equation}

For large spin-splitting, we have 
\begin{equation}
I_{5}\approx \frac{1}{\left( \epsilon^{s}-\epsilon^{-s}\right)
^{2}}\left[ \tau _{s}\delta (\xi^{s})+\tau _{-s}\delta
(\xi^{-s})\right]   \label{q5} \, .
\end{equation}

\section{Spin spiral}
\label{s:spinspiral}

The spin-spiral system has a constant gradient of the rotating
magnetization direction ($a_q =a_{0}\delta _{q,0}$). We perform the local gauge
transformation (\ref{gauge}). The transformed Hamiltonian is
\begin{equation}
\tilde{H}=\frac{\hbar^2}{2m} \sum_{\bf k} {\bf c}_{\bf k}^{\dag
}\left(k^2-\sigma_{z} p^2 +a_{0}k_{z}\sigma
_{y}+\frac{1}{4}a_{0}^{2}\right) {\bf c}_{\bf k} \, ,
\label{Hspiral}
\end{equation}
where ${\bf c}_{{\bf k}}$ is an annihilation operator in the spinor
spin-space and the exchange splitting $\Delta \equiv
\hbar^{2}p^{2}/(2m)$ has been introduced. The transformed current
operator is
\begin{equation}
J=\frac{e\hbar }{m}\sum_{\bf k} {\bf c}_{\bf k}^{\dag }\left(
k_{z}+\frac{a_{0} }{2}\sigma _{y}\right) {\bf c}_{\bf k} \, .
\end{equation}
The Hamiltonian (\ref{Hspiral}) can be exactly diagonalized, and the
eigenvalues are
\begin{equation}
E_{{\bf k}}^{\pm }=\frac{\hbar ^{2}}{2m}\left( k^{2}+\frac{1}{4}a_{0}^{2}\mp
\sqrt{ k_{z}^{2}a_{0}^{2}+p^{4}}\right)
\end{equation}
with the corresponding eigenvectors 
\begin{equation}
{\bf u}_{\pm }={\cal {N}}_{\pm}\left( 
\begin{array}{c}
1 \\ 
i(1\mp \sqrt{1+\alpha ^{2}})/\alpha
\end{array}
\right) \, ,
\end{equation}
where the parameter $\alpha \equiv k_{z}a_{0}/p^{2}$ is
introduced and the normalization factors are
\begin{equation}
{\cal N}_{\pm}^2 =\frac{\alpha^2}{2\sqrt{1+\alpha ^{2}}(\mp
1+\sqrt{1+\alpha ^{2}})}  \, .
\end{equation}
The annihilation operators are transformed as ${\bf c} = \left({\bf
u}_{+},{\bf u}_{-} \right) {\bf a}$. In the new basis, the current
operator is
\begin{equation}
\tilde{J}=\frac{e\hbar }{m}\sum_{\bf k}{\bf a}_{\bf k}^{\dag }\left(
\begin{array}{cc}
k_{z}-\frac{a_{0}\alpha }{2\sqrt{1+\alpha ^{2}}} & \frac{a_{0}}{2\sqrt{
1+\alpha ^{2}}} \\ 
\frac{a_{0}}{2\sqrt{1+\alpha ^{2}}} & k_{z}+\frac{a_{0}\alpha }{2\sqrt{
1+\alpha ^{2}}}
\end{array}
\right) {\bf a}_{\bf k}
\end{equation}
The DC conductivity is
\begin{equation}
\sigma =\frac{1}{4\pi}\frac{\hbar }{V}\sum_{{\bf k}ss' }
\left| J_{ss'} \right| ^{2} A_{{\bf
k}}^{s}A_{{\bf k}}^{s'} \, ,
\label{spiralcond}
\end{equation}
where the electron spectral function ($A=-2\text{Im}G^R$) at the Fermi level is
\begin{equation}
A_{{\bf k}}^{s}=\frac{\hbar
/\tau}{(E_{{\bf k}}^{s}-\mu)^{2}+(\hbar /\tau)^{2}} \, .
\label{spiralspectral}
\end{equation}
Here we have inserted a phenomenological scattering life-time, which
is identical for the two eigenstates.  Note that we cannot treat
different scattering life-times for the minority and majority states
in the bulk ferromagnet with the method outlined in this appendix,
since the life-times appearing in (\ref{spiralcond}) are the
life-times for the exact eigenstates in the spin-spiral. In order to
determine the relation between the different lifetimes, the general
method described above in our paper should be used.  From
(\ref{spiralcond}) and (\ref{spiralspectral}) it can be seen that he
off-diagonal terms in the conductivity, $A_{+}A_{-}$ are in the order
$1/(\tau \Delta /\hbar )^{2}$ smaller than the diagonal terms.  We
further assume that the scattering by the domain wall is weak, {\it
i.e.} $a_{0}^{2}\ll p^{2} $ and $a_{0}k_{F}^s\ll p^{2}$ and expand the
result for the conductivity to the second order in $a_{0}$.  The
conductivity becomes
\begin{equation}
\sigma = \frac{e^{2}\tau}{6\pi^2m}\sum_{s} \left(k_{\mu}^2 - \frac{a_0^2}{4}+s p^2 \right)^{3/2} \left( 1-\frac{sE_w}{4\Delta} \right)  \, ,
\label{spiralc}
\end{equation}
where $\hbar^2 k_{\mu}^2/(2m)=\mu$ and $E_w=\hbar^2 a_0^2/(2m)$ are
used. The conductivity should be related to the electron density by
eliminating any reference to the chemical potential which may change in
the presence of the domain wall,
\begin{equation}
n_{s} = \frac{1}{6 \pi^2} \left(k_\mu^2 -\frac{a_0^2}{4}+s p^2\right)^{3/2} \left(1+ \frac{sE_w}{4\Delta} \right) \, .
\label{spiraldens}
\end{equation}
Inserting (\ref{spiraldens}) into (\ref{spiralc}), the conductivity
can therefore be written as
\begin{equation}
\sigma =\sigma _{0}\left( 1-\frac{n_{+}-n_{-}}{n_{+}+n_{-}}\frac{E_w}{2\Delta}\right) \,,  \label{arne}
\end{equation}
where $\sigma _{0}=e^{2}(n_++n_-) \tau/m$ is the Drude conductivity. The domain
wall resistiviy, $\rho_w=-\delta \sigma_w/\sigma_0^2$ is thus
\begin{equation}
\rho_w = \frac{e^2\rho_0^2 E_w}{2m\Delta} \left(n_+ - n_- \right) \, ,
\end{equation}
where $\rho_0=1/\sigma_0$.

\begin{figure}
\caption{Feynman diagrams of the contributions to the conductivity to
the lowest order in the domain wall scattering. Solid lines indicate
the electron Green's function and the dashed lines the interaction
with the domain wall. The vertex $\times$ arises from the unperturbed
current operator (\ref{cur}), the vertex $\circ$ is due to the gauge
transformation on the current operator (\ref{gaugecur}), the vertex
$\diamond$ is due to the first term in the interaction Hamiltonian
(\ref{V}) and the vertex $\Box$ is due to the second term in the
interaction Hamiltonian (\ref{V}). }
\label{Fdiagram}
\end{figure}

\begin{figure}
\caption{Feynman diagrams of the contributions to the electron density
to the lowest order in the domain wall scattering. Solid lines
indicate the electron Green's function and the dashed lines the
interaction with the domain wall. The vertex $\diamond$ is due to the
first term in the interaction Hamiltonian (\ref{V}) and the vertex
$\Box$ is due to the second term in the interaction Hamiltonian
(\ref{V}).}
\label{Fdensity} 
\end{figure}

\begin{figure}
\caption{The relative change in the resistiviy, $\rho_w/(\kappa \rho_0)$,
due to the scattering by the domain wall as a function of the
asymmetry in the scattering life-times $\tau_+/\tau_-$. The solid line
is for $\gamma=k_{F}^+/k_{F}^-=1.01$, the dashed line is for
$\gamma=k_{F}^+/k_{F}^-=1.20$ and the dotted line is for
$\gamma=k_{F}^+/k_F^-=10.0$.}
\label{f:rho}
\end{figure}


\begin{references}

\bibitem[\dag]{Brataas} also at Philips Research Laboratories, Prof.\
Holstlaan 4, 5656 AA Eindhoven, The Netherlands.

\bibitem{Gregg96:1580} J.F. Gregg, W. Allen, K. Ounadjela, M. Viret,
M. Hehn, S. M. Thompson and J. M. D. Coey, Phys. Rev. Lett. {\bf 77},
1580 (1996) ; M. Viret, D. Vignoles, D. Cole, J. M. D. Coey, W. Allen,
D. S. Allen and J. F. Gregg, Phys. Rev. B {\bf 53}, 8464 (1996).

\bibitem{Hong98:L401}  K. M. Hong, N. Giordano, J. Phys.: Condensed Matter, 
{\bf 10}, L401 (1998).

\bibitem{Otani98:1096} Y. Otani, S. G. Kim, K. Fukamichi, O. Kitakami
and Y. Shimada, IEEE Trans. Magn., {\bf 34}, 1096 (1998).

\bibitem{Ruediger98:5639} A. D. Kent, U. Ruediger, J. Yu, S. Zhang,
P. M. Levy, Y. Zhong and S. S. P. Parkin, IEEE Trans. Magn., {\bf 34},
900 (1998); U. Ruediger, J. Yu, S. Zhang, A, D. Kent, S. S. P. Parkin,
Phys. Rev. Lett., {\bf 80}, 5639 (1998); A. D. Kent, U. Ruediger,
J. Yu, L. Thomas and S. S. P. Parkin cond-mat/9812163.

\bibitem{Gorkom99:422} R. P. van Gorkom, J. Caro, S. J. C. H. Theeuwen,
K. P. Wellock, N. N. Gribov, Appl. Phys. Lett. {\bf 74}, 422 (1999).

\bibitem{Tatara97:3773}  G. Tatara and H. Fukuyama, Phys. Rev. Lett. {\bf 67}
, 3773 (1997).

\bibitem{Levy97:5110}  P.M. Levy and S. Zhang, Phys. Rev. Lett. {\bf 79},
5110 (1997)

\bibitem{vanHoof99:138} J. van Hoof, K. M. Schep, A. Brataas,
G. E. W. Bauer and P. J. Kelly Phys. Rev. B {\bf 59}, 138 (1999).

\bibitem{Garcia99:2923} N. Garcia, M. Mu\~{n}oz, and Y.-W. Zhao,
Phys. Rev. Lett. {\bf 82}, 2923 (1999).

\bibitem{Cabrera74:217}  G.G. Cabrera and L.M. Falicov, Phys. Stat. Sol. B 
{\bf 62}, 217 (1974); {\bf 61}, 539 (1974).

\bibitem{Berger84:1954} L. Berger, J. Appl. Phys. {\bf 55}, 1954
(1984).

\bibitem{Ruediger98}  U. Ruediger, J. Yu, S. S. P. Parkin and A. D. Kent, in
press.

\bibitem{Ruediger9901245} U. Ruediger, J. Yu, L. Thomas,
S. S. P. Parkin and A. D. Kent cond-mat/9901245.


\bibitem{Brataas98:545}
A. Brataas, G. Tatara and G. E. W. Bauer, Phil. Mag. B {\bf 78}, 545 (1998).


\bibitem{Mertig94:11767}  I. Mertig, R. Zeller and P. H. Dederichs, Phys.
Rev. B {\bf 49}, 11767 (1994); P. Zahn, I. Mertig, M. Richter and H. Eshrig,
Phys. Rev. Lett. {\bf 75}, 2996 (1995).

\bibitem{Bender78} C. M. Bender and S. A. Orszag, {\it Advanced
Mathematical Methods for Scientists and Engineers} (McGraw-Hill, New
York, 1978).
\end{references}
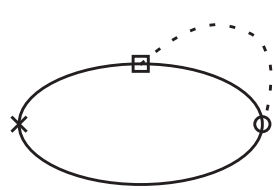
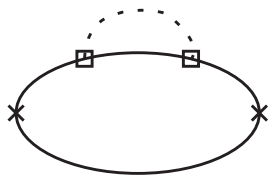
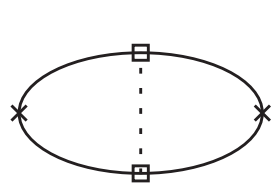
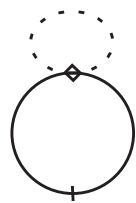
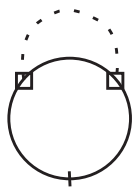
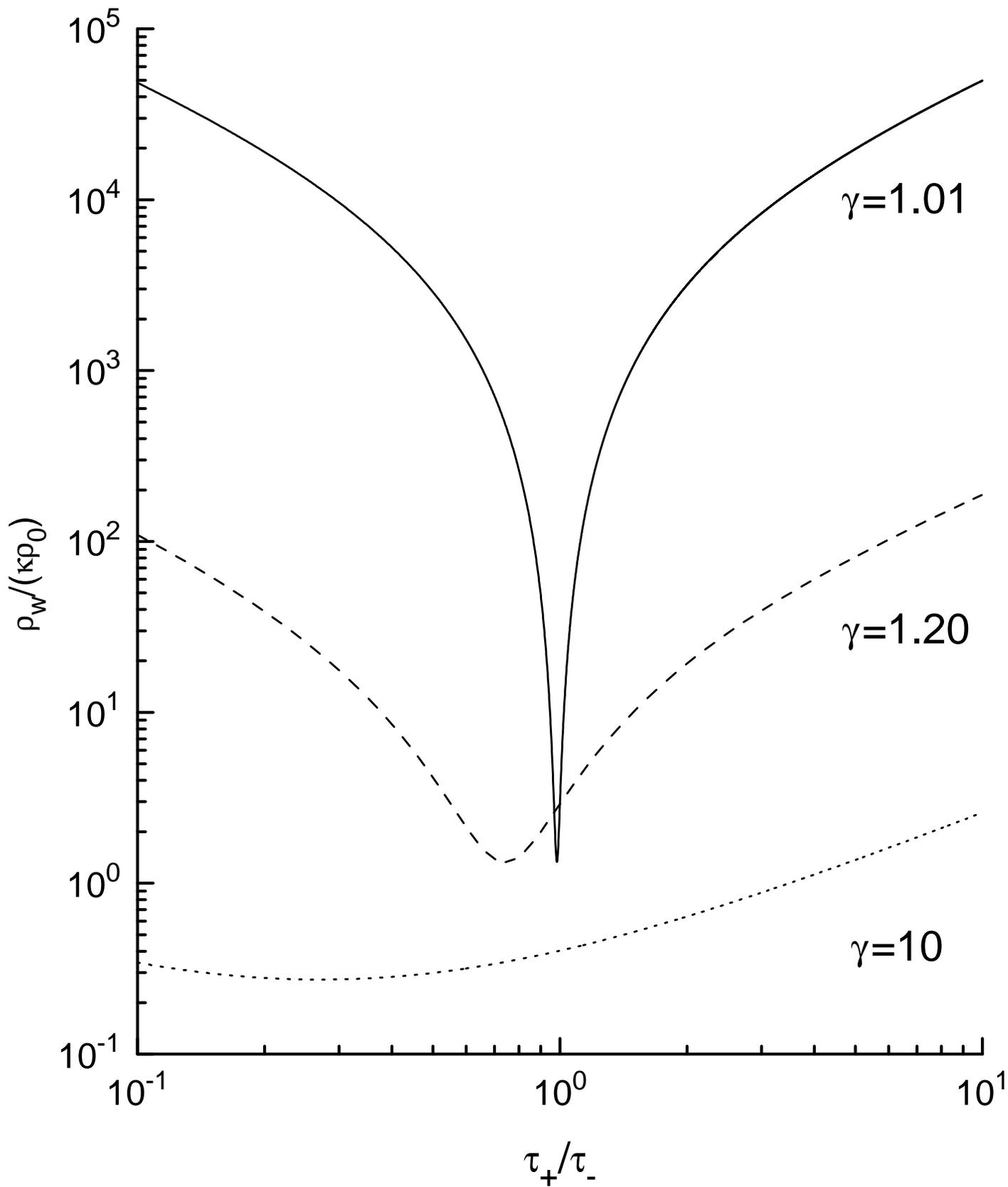
\end{document}